\documentclass[preprint,authoryear,12pt]{elsarticle}
\usepackage{graphicx}
\usepackage{color, ulem, upgreek, microtype}
\usepackage{amssymb,lineno}
\usepackage[
breaklinks=true,%
colorlinks=true,%
pdfauthor={First Author et al.},%
pdftitle={Template  for manuscripts in Advances in Space Research}%
]{hyperref}

\journal{Advances in Space Research}
\begin{document}
\newpage
\pagenumbering{alph}\setcounter{page}{1} 
\setpagewiselinenumbers
\modulolinenumbers[2]
\newpage
\pagenumbering{arabic}\setcounter{page}{1}
\begin{frontmatter}


\title{Internal wave coupling processes in Earth's atmosphere}


\author[gmu]{Erdal Yi\u git\corref{cor}}
\address[gmu]{George Mason University, Fairfax, Virginia, USA}
\cortext[cor]{Corresponding author}
\ead{eyigit@gmu.edu}

\author[mps,goe]{\rm Alexander S. Medvedev}
\address[mps]{Max Planck Institute for Solar System Research,
G\"ottingen, Germany}
\address[goe]{Institute of Astrophysics, Georg-August University G\"ottingen,
Germany\\~\\
  {\rm \scriptsize \bf This review paper has been accepted for publication in Advances in Space
  Research.}
}
\ead{medvedev@mgs.mps.de}

\begin{abstract}
  This paper presents a contemporary review of vertical coupling in the
  atmosphere and ionosphere system induced by internal waves of lower
  atmospheric origin. Atmospheric waves are primarily generated by
  meteorological processes, possess a broad range of spatial and temporal
  scales, and can propagate to the upper atmosphere.  A brief summary of
  internal wave theory is given, focusing on gravity waves, solar tides,
  planetary Rossby and Kelvin waves. Observations of wave signatures in the
  upper atmosphere, their relationship with the direct propagation of waves into
  the upper atmosphere, dynamical and thermal impacts as well as concepts,
  approaches, and numerical modeling techniques are outlined. Recent
  progress in studies of sudden stratospheric warming and upper atmospheric
  variability are discussed in the context of wave-induced vertical coupling
  between the lower and upper atmosphere.
\end{abstract}


\end{frontmatter}

\parindent=0.5 cm

\section{Introduction to Atmospheric Vertical Coupling}
\label{sec:intro}

The structure and dynamics of the Earth's atmosphere are determined by a complex
interplay of radiative, dynamical, thermal, chemical, and electrodynamical
processes in the presence of solar and geomagnetic activity variations. The
lower atmospheric processes are the primary concern of meteorology, while
impacts of the Sun and geomagnetic processes on the atmosphere-ionosphere are
the subject of space weather research. Thus, the whole atmosphere system is
under the continuous influence of meteorological effects and space weather.  A
detailed understanding of the coupling mechanisms within the
atmosphere-ionosphere is crucial for better interpreting atmospheric
observations, understanding the Earth climate system, and developing forecast
capabilities.

The atmosphere can be viewed as an ideal geophysical fluid that is pervaded by
waves of various spatio-temporal scales. The term ``internal" signifies the
ability of waves to propagate ``internally", that is, vertically upward within
the atmosphere, unlike ``external" modes, in which all layers oscillate in sync
when disturbances propagate
horizontally. Internal waves exist because Earth's atmosphere is overall stably
stratified. Horizontal scales of internal waves vary from a few kilometers to
the planetary circumference. Temporal scales cover a range from minutes to
several days. Internal waves can propagate over large distances, and transfer
momentum and energy from lower levels to much higher altitudes, thus providing
an important coupling mechanism in the atmosphere. What are the impacts of these
waves on the atmosphere-ionosphere system at various scales?  The scientific
community has increasingly been realizing that answering this
question represents a crucial step toward better understanding the connections
between meteorology and space weather.

Coupling processes from the lower atmosphere to the ionosphere have been the
overarching goal of recent observational campaigns.  One such campaign was the
SpreadFEx that was conducted over the South American sector \citep{Abdu_etal09,
  Takahashi_etal09}.

Nonlinear interactions of internal waves between themselves and with 
the undisturbed atmospheric flow create a complex dynamical system, in
which long-range coupling processes can occur that link different 
atmospheric layers and redistribute energy and momentum between them. 
To investigate the origins and global consequences of such processes, 
the atmospheric layers cannot be investigated in isolation, but the 
atmosphere ought to be treated as a whole system. 
The importance of such investigations have been broadly recognized. 
In the Role Of the Sun and the Middle atmosphere/thermosphere/ionosphere 
In Climate \citep[ROSMIC,][]{Lubken_etal14} project within the 
SCOSTEP's VarSITI (Variability of the Sun and Its Terrestrial Impact) 
program, the influence of the lower atmosphere on the upper
atmosphere is designated as one of the focus topics \citep{Ward_etal14}. 
Specifically, ROSMIC's ``Coupling by Dynamics" Working Group 
coordinates the efforts in investigating dynamically-induced 
vertical coupling processes in the atmosphere-ionosphere system.

\begin{figure}\centering
  \includegraphics*[width=1.\textwidth,angle=0]{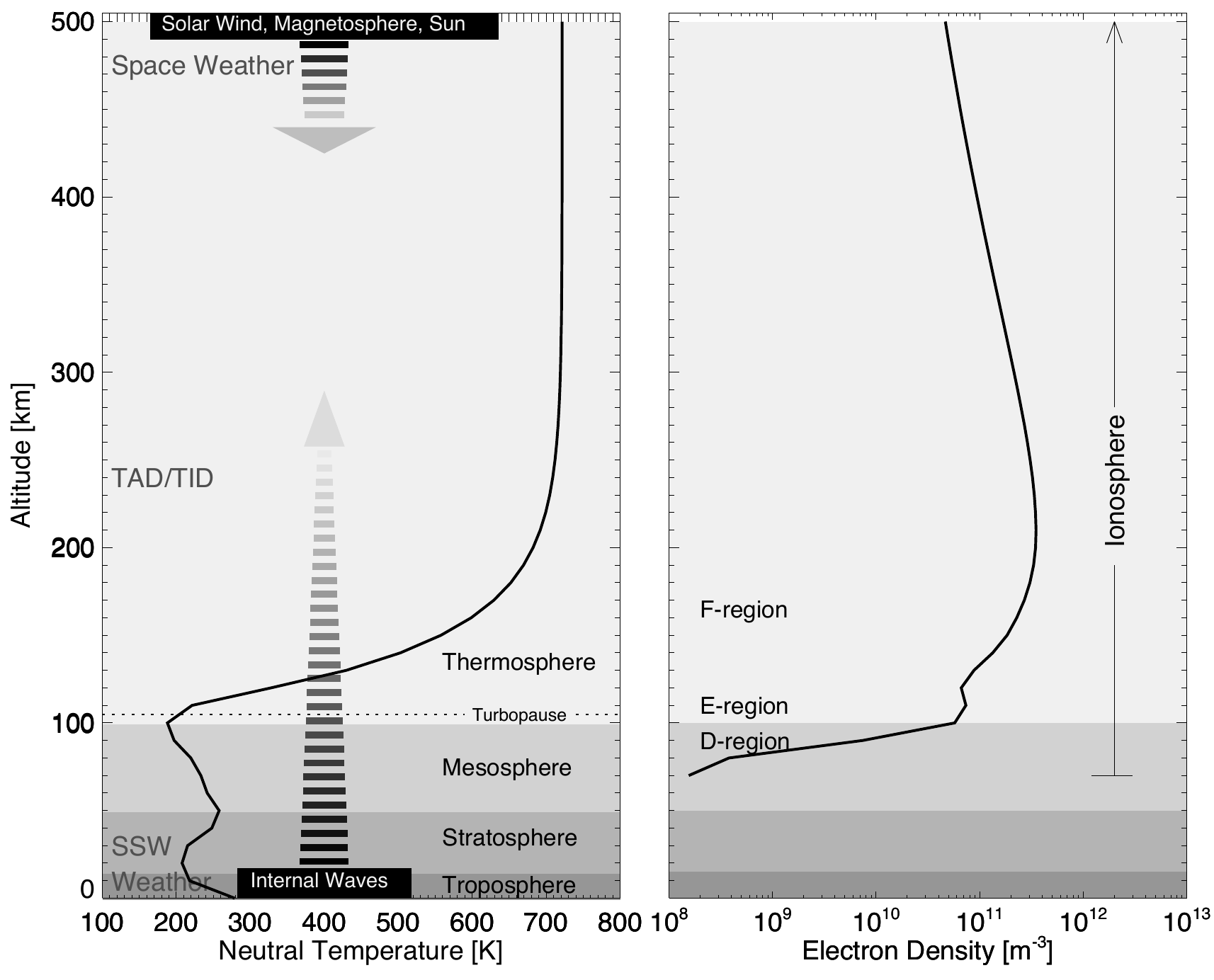}
  \caption{Vertical structure of the atmosphere-ionosphere system, where the
    neutral atmospheric temperature is shown on the left and the electron
    density distribution on the right. Panels are produced using midlatitude
    data from  MSISE-90 and IRI 2012 models for 1 January 2010 at noon with
    daily $F_{10.7}=77.2\times 10^{-22}~\rm{W}~\rm{m}^{-2}~\rm{Hz}^{-1}$ and
    $A_p=0.5$. SSW and TAD/TID denotes sudden stratopspheric warming and
    traveling atmospheric/ionospheric disturbances, respectively. The turbopause
    is at about 105 km.}
  \label{fig:atmion}
\end{figure}

Figure \ref{fig:atmion} illustrates the different regions in the
atmosphere-ionosphere system taken from the empirical models of MSISE-90 and
IRI-2012 \citep{Bilitza_etal14}. The characteristic distribution of the neutral
temperature with height determines the major regions in the neutral atmosphere
shown on the left: the troposphere, stratosphere, mesosphere, and thermosphere,
where the latter is the hottest region of the neutral atmosphere. The ionosphere
is characterized by the electron density profile and represents an ionized
portion that is produced largely by solar irradiation, and coexists with the
neutral upper atmosphere. The main ionospheric regions are D-, E, and F-regions,
where the peak electron density is found within the F-region. The physical
processes that influence the atmosphere-ionosphere system from below
(``internal waves") and above (``solar wind, magnetosphere, Sun") are
shown in black boxes as well. Internal waves are associated with the
lower atmospheric weather (i.e., meteorological effects), and the effects
caused by solar wind, magnetosphere, and Sun are termed broadly as space
weather. Various atmosphere-ionosphere transient events, such as sudden
stratospheric warming (SSW) and traveling atmospheric/ionospheric disturbances
(TAD/TID) are depicted in gray at approximate altitudes where they
typically occur. The turbopause ($\sim$105 km) marks the hypothetical
boundary between the turbulently mixed homosphere and the heterosphere, where
diffusive separation dominates.

A number of reviews on various aspects of coupling processes in the
atmosphere-ionosphere system have been presented previously by various authors
\citep[e.g.,][]{Kazimirovsky_etal03, Altadill_etal04, Lastovicka06, Forbes07,
  Lastovicka09a, Lastovicka09b}. The observations of long-term trends in the
upper atmosphere have been brought to the attention of the scientific community
in the works of \citet{Lastovicka12} and \citet{Lastovicka13}.  Here we focus on
the role of internal waves in the coupling between the lower and upper
atmosphere, and the resulting larger-scale effects.  In particular, sudden
stratospheric warmings and wave-induced
atmospheric variability are discussed in the context of vertical coupling, as
the two topics have recently come again to the meteorology and aeronomy
communities' attention. Our overarching goal is to provide a motivation for
bridging gaps between scientists studying the lower, middle, and upper
atmospheres. In this paper, we concentrate on the most recent studies, while
providing references to the existing reviews for further details.

This review is structured in the following manner.  Section \ref{sec:wave}
provides an overview of the physical properties of internal waves. Section
\ref{sec:obs} outlines the observations of wave structures in the middle and
upper atmosphere.  Section \ref{sec:model} presents modeling techniques for
studying internal waves, and section~\ref{sec:effect} gives some details
concerning atmospheric wave propagation and consequences for the upper
atmosphere. Sections \ref{sec:ssw} and \ref{sec:uav} discuss coupling processes
during sudden stratospheric warmings (SSWs), and other effects related to upper
atmosphere variability.  In section \ref{sec:conc}, conclusions are given, and
some open questions are highlighted.

\section{Internal Wave Characteristics and Propagation}
\label{sec:wave}
To a first approximation, atmospheric waves can be distinguished by their
spatial scales.  Earth's atmosphere possesses a broad spectrum of waves ranging
from very small- (e.g., gravity waves, GWs) to planetary-scale waves (tides,
Rossby waves). Table \ref{tb:table1} summarizes quantitatively the range of
temporal scales for tides, gravity, planetary Rossby and Kelvin waves. Overall,
internal wave periods vary from few minutes to tens of days. They also have
different spatial scales.  While small-scale GWs have typical horizontal
wavelengths $\lambda_H$ of several km to several hundred km, horizontal scales
of solar tides and planetary waves are comparable to the circumference of Earth.

Although various internal waves can be excited by different mechanisms, in
general, meteorological processes are the primary sources of these motions. They
can propagate upward and grow in amplitude due to exponentially decreasing
neutral mass density $\rho$ (in order to satisfy wave action conservation).
Therefore, although wave disturbances associated with small-scale GWs are
  relatively small at the source levels, their amplitudes can become
significant at higher altitudes in the thermosphere, and are all subject to
various dissipation processes. On the other hand, large-scale waves, such as
  planetary waves, can possess relatively larger amplitudes in the lower
  atmosphere, and can, therefore, dissipate at lower altitudes. This wave
dissipation is the mechanism of transfer of momentum and energy from
disturbances to the mean flow. Below, a brief characterization of some internal
waves is presented.

\subsection{Gravity Waves}
\label{sec:2-GW}
These waves have a broad range of scales from small-scale acoustic-gravity waves
to large-scale inertia-gravity waves.  They are routinely seen in lidar, radar,
airglow, and satellite measurements. Generated typically in the lower
atmosphere by meteorological processes, such as convection \citep{Song_etal03},
frontogenesis \citep{Gall_etal88}, nonlinear interactions
\citep{MedvedevGavrilov95}, and thunderstorms \citep{CurryMurty74}, they can
propagate upward to the middle and upper atmosphere.  These sources produce
broad spatial and temporal spectra of GWs in the lower atmosphere.  The spectra
typically describe quadratic (with respect to disturbances) quantities: energy,
wave action, or momentum fluxes ($\overline{u^\prime w^\prime}$,
$\overline{v^\prime w^\prime}$) as functions of horizontal phase speeds $c$, or
other spectral parameters. The fluxes $\overline{u^\prime w^\prime}$ and
$\overline{v^\prime w^\prime}$ denote the vertical fluxes of the wave zonal and
meridional momentum, respectively. The vertical structure of fluxes associated
with a GW harmonic $i$ can be written as \citep[e.g.,][]{Yigit_etal08}:
\begin{eqnarray}
  \textrm{\bf F} = \left\{ \begin{array}{l}
    \overline{u^\prime w^\prime_i(z)}  \\
    \overline{v^\prime w^\prime_i(z)}    
  \end{array} \right\} = 
  \left\{ \begin{array}{l}
    \overline{u^\prime w^\prime_i(z_0)}  \\
    \overline{v^\prime w^\prime_i(z_0)}  
  \end{array} \right\}\cdot
  \frac{\rho(z_0)}{\rho(z)}\tau_i(z),
\label{eq:flux}
\end{eqnarray}
where $z_0$ denotes the source level; $\rho(z) = \rho(z_0)\exp[-(z-z_0)/H]$ is
the background neutral mass density, $H$ is the scale height, and $\tau_i$ is
the transmissivity function for the $i^{\rm th}$ harmonic.  Equation
(\ref{eq:flux}) implies that the wave momentum flux grows exponentially with
height $z$ above the source level $z_0$.  As GWs propagate upward, they interact
with the atmospheric background continuously. The mean flow affects the
propagation of GWs by modifying the transmissivity, which is given for one
harmonic by
\begin{equation}
  \tau_i(z) = \exp \Bigg[  -\int_{z_0}^z\sum_d \beta_d^i(z^\prime)dz^\prime  \Bigg],
\label{eq:tau}
\end{equation}
where $\beta_d^i$ is the dissipation of wave harmonic $i$ due to a given 
attenuation process denoted by $d$. The total dissipation is the 
sum of all attenuation processes taking place simultaneously:
\begin{equation}
  \beta_{tot}(z) = \beta_{ion}(z) + \beta_{mol}(z) + \beta_{eddy}(z) + \beta_{non}(z) +
  \beta_{rad}(z),
  \label{eq:beta}
\end{equation}
where the height-dependent total dissipation $\beta_{tot}(z)$ is a sum of 
dissipations due to ion drag, molecular viscosity and thermal conduction 
\citep{VadasFritts05}, eddy viscosity, nonlinear diffusion
\citep{Weinstock82, MedvedevKlaassen95, MedvedevKlaassen00}, and
radiative damping \citep{Holton82}, as introduced in the work by
\citet{Yigit_etal08}. 

As seen from (\ref{eq:flux})--(\ref{eq:beta}), upward GW 
propagation is affected by a competition between the nearly 
exponential growth of the momentum flux $F$, and dissipation 
acting upon the wave. This process facilitates a continuous 
transfer of the wave momentum and energy to the mean flow. 
The resulting dynamical effect ${\bf a}$ on the flow represents a 
body force per unit mass, that is, the acceleration/deceleration 
of the mean flow, which is given by the divergence of the 
momentum flux ${\rm \bf F}$:
\begin{equation}
  {\bf a}(z) = -\frac{1}{\rho} \frac{\partial \rho {\rm\bf F}(z)}{\partial z}.
\end{equation}
Thermal effects originating from GWs are described by 
heating/cooling rates per unit mass. The total thermal effect
comprises an irreversible heating $E$ due to transfer of
mechanical energy into heat, and the differential heating/cooling 
$Q$ caused by the downward transport of 
the sensible heat flux $\overline{w^\prime T^\prime}$:
\begin{equation}
Q(z) = -\frac{1}{\rho} \frac{\partial \rho \overline{w^\prime T^\prime}}
     {\partial z},
\end{equation}
$T^\prime$ being wave-induced temperature disturbances
\citep{MedvedevKlaassen03,Becker04,YigitMedvedev09}.
The very same formalism for calculating dynamical and thermal effects
can be applied to solar tides, planetary and Kelvin waves described below.

\subsection{Solar Tides}
\label{sec:tide}
Solar tides are particular types of gravity waves forced by periodic heating
associated with the absorption of solar radiation in the atmosphere.  The
obvious period is 24 hours, however, due to the nonlinearity of the atmospheric
dynamics, higher-order harmonics with frequences $\omega_n=n\Omega$,
$\Omega=2\pi /24{\rm \enskip hours}$ ($n=2, 3$, etc.)  are also excited. The
first few tidal harmonics (diurnal, semidiurnal, and terdiurnal) are abundant in
the middle atmosphere, and have noticeable amplitudes. Tidal disturbances can be
represented globally with the Fourier series as a sum of harmonics
\begin{equation}
  A_{ns}\cos\,(n\Omega t + s\lambda +\phi_{ns}),
  \label{eq:tide1}
\end{equation}
where $\lambda$ is the longitude, $s=0, \pm 1, \pm2, ...$ is the zonal
wavenumber, and $A_{ns}$ and $\phi_{ns}$ are the amplitude and phase,
correspondingly. Eastward and westward propagation correspond to $s<0$ and
$s>0$, respectively. Equation (\ref{eq:tide1}) can be rewritten in terms of the
local time $t_{LT}=t+\lambda/\Omega$ as
\begin{equation}
  A_{ns}\cos\,[n\Omega t_{LT} + (s-n)\lambda + \phi_{ns}].
  \label{eq:tide2}
\end{equation}
The nomenclature of tides follows from (\ref{eq:tide2}). If $s=n$, oscillations
are Sun-synchronous (that is, waves move westward following the apparent motion
of the Sun), and such modes are called ``migrating tides". Harmonics with $s\ne
n$ are non-migrating tides, which can move either westward or eastward
(depending on $s$ and $n$). Such harmonics are excited due to inhomogeneity of
solar radiation absorption and nonlinearity of atmospheric motions.  

Tides have distinctive meridional structures defined by their propagation
  properties. Under the approximation of a linear, motionless, isothermal, and
  inviscid atmosphere, the horizonal structure of tides can be conveniently
  represented as a sum of the so-called Hough functions comprising symmetric and
  antisymmetric with respect to the equator modes.  In the real atmosphere, the
  fundamental tidal structure no longer coincides with the Hough functions. In
  a practical sense, however, they are still sometimes used in
  data analyses, because only a few Hough modes can provide a good fit for the
  tidal component.

As with all gravity waves, only tides, whose intrinsic frequency
$\hat{\omega}=\omega_n- s\bar{u}(a\cos\phi)^{-1}$ ($a$ being the Earth radius,
$\phi$ is the latitude, $\bar{u}$ is the zonally averaged wind) exceeds the
local Coriolis frequency $f=2\Omega \sin\phi$, can propagate vertically and
produce an effective coupling.  For instance, the lower-frequency diurnal tide
can propagate from tropospheric heights into the upper atmosphere mainly at
low-latitudes (subject to the local winds $\bar{u}$), and is vertically trapped
and rapidly decays above the excitation levels at higher latitudes.  Tides are
observed in the middle atmosphere and thermosphere, where they play a profound
role in maintaining atmospheric variability.  At the heights of the F region,
strong \textit{in situ} thermal tides are forced by the periodic heating due to
absorption of solar UV and EUV radiation. These tides have a barotropic
structure (no phase shift in vertical), and the diurnal mode dominates. More on
the importance of tides in the upper atmosphere can be found in the review paper
of \citet{Forbes07}.

\subsection{Planetary Waves}
\label{sec:pw}
Planetary waves are often a synonym of Rossby waves, which appear in the
atmosphere due to the latitudinal gradient of the Coriolis force that balances
variations of the pressure gradient force. Their phase speed is always westward,
however the group velocity can have either direction. Planetary waves manifest
themselves as a meandering of jet streams, where the number of meanders gives
the zonal wavenumber.  Rossby waves are strongly dispersive: those having faster
speeds are usually trapped and do not propagate vertically. They are called
``barotropic" modes, unlike the slower moving ``baroclinic" harmonics with zonal
wave speeds of few cm~s$^{-1}$. They are excited by the instability of the
troposopheric jet: barotropic (due to a horizontal wind shear) or baroclinic one
(due to a vertical wind shear), correspondingly. The slowest planetary waves
with the horizontal wavenumber 1 and/or 2, which are locked to particular
locations, are called ``quasi-stationary" waves, and are forced by surface
inhomogeneities -- topography and/or surface temperature. According to the
dispersion relation for Rossby waves, which involves the background zonal wind
velocity, the quasi-stationary waves have higher chances of
propagating into the middle atmosphere and depositing their momentum and energy
upon dissipation, mostly in the winter hemisphere 
Rossby waves occur at middle- and high-latitudes, although
they sometimes can shift to lower latitudes, and even cross the equator
\citep{Medvedev_etal92}.  Planetary waves play a profound role in the dynamics
of the stratosphere.  In particular, the wave momentum they deposit drives the
pole-to-pole circulation. Another well-known transient phenomenon associated
with planetary waves are sudden stratospheric warmings (see section
\ref{sec:ssw}). Higher in the atmosphere, the dynamical importance of Rossby
waves reduces leaving it to gravity waves.

\subsection{Kelvin Waves}
\label{sec:kelvin}
Atmospheric Kelvin waves are large-scale waves that are the results of
balancing the Coriolis force against the waveguide at the equator. 
These waves are trapped at low latitudes, their amplitudes decay 
steeply away from the equator, and their meridional velocity component 
vanishes, while the zonal one fluctuates. An important feature of Kelvin 
waves is that they are
non-dispersive, that is, the phase speed of the wave crests is equal to 
the group velocity at all frequencies. The second peculiarity of 
atmospheric Kelvin waves is that their phase velocity is positive. 
Thus, Kelvin wave disturbances always propagate eastward retaining their 
shapes, which is a great help in detecting them.

Kelvin waves are excited by convection in the lower atmosphere, and can
propagate vertically. As they dissipate in upper layers, the westerly momentum
they deposit affects the mean circulation. Kelvin wave classification includes
slow (periods of 10 to 20 days), fast (6 to 10 days), and ultra-fast (3 to 5
days). Slow waves play an important role in the lower stratosphere dynamics,
e.g., in forcing the quasi-biennial oscillation, fast are noticeable in the
entire stratosphere, while ultra-fast harmonics can reach the mesosphere and
thermosphere heights, providing the coupling of the latter with the equatorial
lower atmosphere. A good review of recent observations and effects of Kelvin
waves in the middle and upper atmosphere can be found in the introduction of the
paper by \citet{DasPan13}.

\section{Observations of Internal Wave Activity in the Middle and Upper 
Atmosphere}
\label{sec:obs}
\subsection{General Characteristics of Observations}
A variety of techniques is utilized to detect wave signatures in the 
atmosphere. To date, a combination of \textit{in-situ} and remote 
sensing observational methods provides an unprecedented view of the 
local and global state and composition of the atmosphere. 
Thus, atmospheric temperature, pressure, wind fields, humidity, 
solar radiation flux, trace substances, electrical properties, and 
precipitation can be observed by these techniques.
A variety of small- and large-scale structures can be identified in 
the measured fields, which exhibit systematic wave-like variations and, 
thus, are indicative of wave propagation. As more technologically 
sophisticated instruments with higher resolutions are built, the 
capability of capturing finer structures increases. 

The key general characteristics of observations are error,
accuracy, resolution, and signal-to-noise ratio. An error describes the 
statistical deviation of a measurement from the true value. 
An estimate of the precision of a measurement is 
the root mean square error. Signal-to-noise ratio is the ratio 
between the measured value and the background and instrument noise. 

An \textit{in-situ} measurement means that the sensor is directly in contact
with the matter whose properties are to be measured, implying a certain level of
interaction between the matter and the device that can affect measurements. In
remote sensing, there is no direct contact between the instrument and the
matter.  Surface-based and in-situ observations provide only a very limited
global coverage, while remote sensing methods do not have this shortcoming.
They can be conducted via facilities on the ground, aircraft, and on satellites.
Active sounding remote sensing techniques, such as radar and lidar, emit a beam
with a known intensity and wavelength, and analyze the backscattered
signal. Satellites that can incorporate multiple instruments on board provide
the largest and most comprehensive global coverage, while their accuracy
and temporal resolution may be less than those of \textit{in-situ} ones.  The
first meteorological satellite TIROS (Television and InfraRed Observation
Satellite) was launched in 1960, and enabled the first sounding of the
terrestrial atmosphere from space.  Nowadays, a large number of orbiters is
operated around geospace.

\subsection{Internal Wave Observations}

\begin{figure}\centering
    \includegraphics*[width=0.8\textwidth]{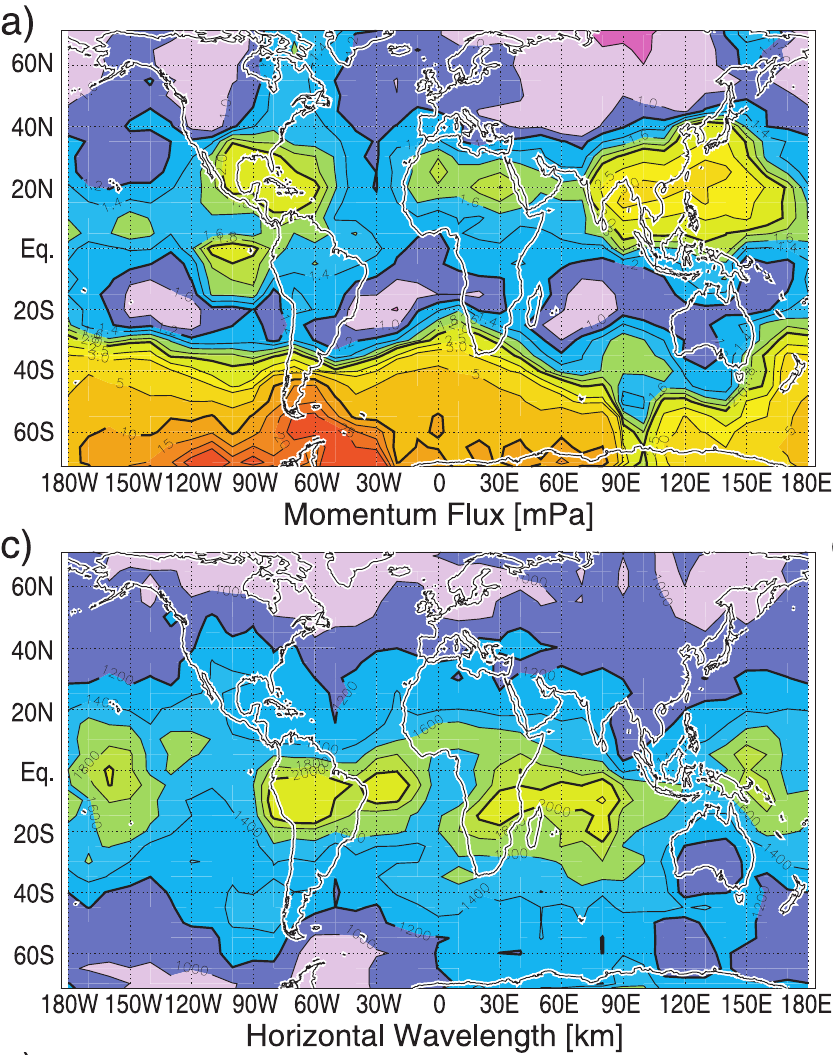}
    \includegraphics*[width=0.8\textwidth]{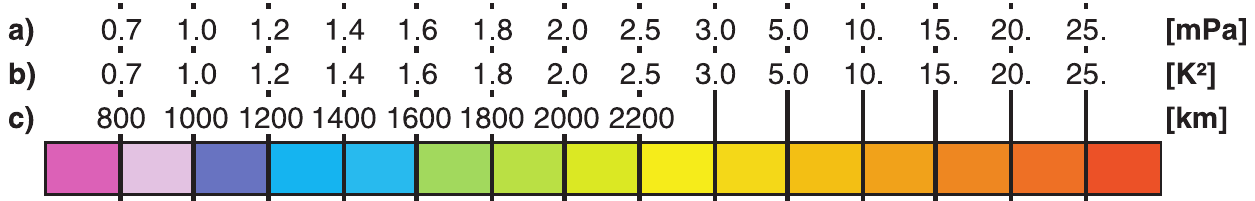}
  \caption{Absolute values of the vertical flux of horizontal gravity wave
    momentum in mPa (above) and average horizontal wavelengths (below) in km
    observed by CRISTA-2 (Cryogenic Infrared Spectrometers and Telescopes for
    the Atmosphere-2) at 25 km in August 1997. Adopted from Figure 3 of
    \citet{Ern_etal04}.}
  \label{fig:gwobss}
\end{figure}

Satellite observations can be utilized for characterization of gravity waves
both in the lower and upper atmosphere.  Figure~\ref{fig:gwobss} shows the
global distribution of GW activity taken from Figure~3 of
\citet{Ern_etal04}. The upper panel gives the latitude-longitude distribution of
the absolute values of GW horizontal momentum (in mPa), and the lower panel
presents the average horizontal wavelengths at 25 km in June 1997 determined by
the Cryogenic Infrared Spectrometers and Telescopes for the Atmosphere
\citep[CRISTA,][]{Offermann_etal99} satellite. It is seen that large fluxes are
concentrated in the winter hemisphere and at Northern Hemisphere low-latitudes.
Observed GWs have larger average horizontal scales in the tropics than at
middle- and high-latitudes. This instrument could capture waves with horizontal
scales $\lambda_H$ greater than 100 km.

\begin{figure}\centering
    \includegraphics*[width=0.8\textwidth]{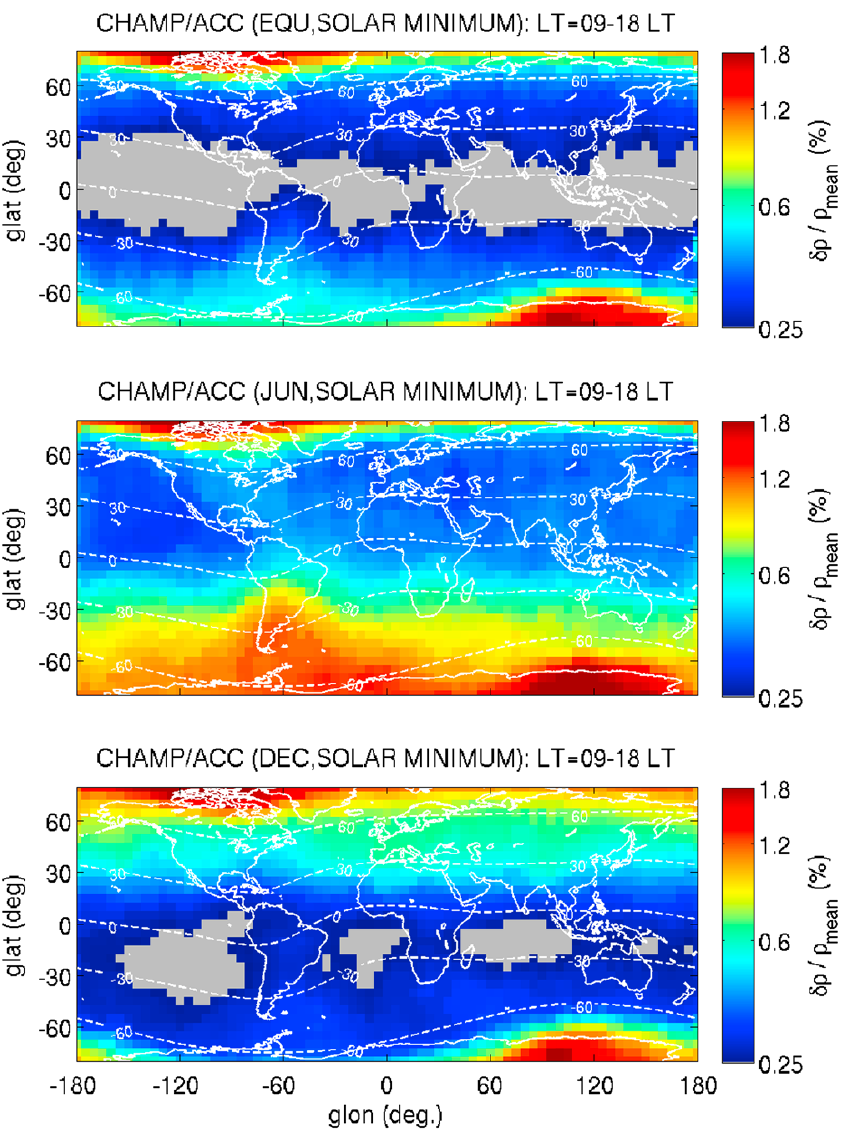}
    \caption{Day-time gravity wave activity in terms of relative density
      perturbations during equinox, June and December solstices based on solar
      minimum years (2006--2010) obtained from the CHAMP
        satellite \citep[Figure 2,][]{Park_etal14}.}
  \label{fig:gwobs}
\end{figure}

Another example of satellite observations is given in
Figure~\ref{fig:gwobs}, which shows GW activity in the 
thermosphere at 400 km retrieved from the CHAMP (Challenging 
Minisatellite Payload) satellite \citep{Park_etal14} launched in
2000. Latitude-longitude distributions of the daytime relative 
density perturbations induced by GWs are shown for equinox, June 
and December solstices at low solar activity. These authors have 
also constructed a map of GW activity in the thermosphere 
at high solar activity, and found that the gravity wave 
fields exhibit similar morphologies, but the amplitudes are a 
factor of two weaker.

Gravity waves produce fluctuations in the atmospheric airglow 
intensity \citep{Hickey_etal93} that can be observed by imaging techniques
\citep{Taylor97}. \citet{Mende_etal98} have conducted satellite-born 
observations of 762 nm O$_2$ airglow. \citet{Frey_etal00} have measured 
GW-induced fluctuations of the hydroxyl OH band. More recently, 
\citet{Tang_etal14} observed high-frequency GW characteristics with an 
all-sky OH airglow imager based on the data from 2003--2009. They have
found typical horizontal scales of 20--40 km, and phase speeds of 30--70 
m~s$^{-1}$. Incoherent scatter radar measurements suggest that gravity 
wave-like variations can be detected frequently in the upper atmosphere 
\citep{Oliver_etal97, Djuth_etal04, Livneh_etal07}.

Unlike with highly irregular GW fields, tidal characteristics 
are easier to obtain due to their periodicity and global-scale sizes.
Tides can be observed both with satellite and ground-based techniques.  
Time and height variations of temperature and wind changes caused by 
tides can be measured continuously by incoherent scatter radars at 
single sites. Ground-based observations have provided information on 
the seasonal and interannual variations of the diurnal and semidiurnal 
modes \citep[e.g.,][]{Yuan_etal06, Fritts_etal10}. Despite
some limitation in coverage, a global picture of tidal activity can be 
obtained from multiple sites \citep{Pancheva_etal02}.

Tidal variability can be inferred from satellite measurements of 
infrared brightness temperature \citep{HaganForbes02}. For a global 
coverage, satellite measurements of the Doppler-shift of airglow at 
different heights in the thermosphere have been conducted by the Wind 
Imaging Interferometer (WINDII) instrument on board UARS (Upper 
Atmosphere Research Satellite). 

In the mesosphere and lower thermosphere (MLT), migrating and 
nonmigrating solar tidal components in temperature and neutral winds 
have frequently been observed using SABER and TIDI instruments on 
board the Thermosphere-Ionosphere-Mesosphere Energetics and Dynamics
(TIMED) satellite \citep{Killeen_etal06, Wu_etal06}, and the two 
components could be separated from each other \citep{Oberheide_etal05,
Pancheva_etal13}. Usually, monthly mean tidal amplitudes and phases are 
derived from the observed temperature and zonal and meridional winds, and 
density residuals \citep{OberheideForbes08a, OberheideForbes08b}. 
In order to derive tidal signatures from satellite data, a 24-hour 
local time coverage is required. Along the orbit of the satellite, 
the longitude and the universal time vary simultaneously. Local time 
precession of the satellite was used to provide a full local time
coverage and derive tidal signatures. Using density residuals derived 
from accelerometers on board CHAMP and 
Gravity Recovery and Climate Experiment (GRACE) 
satellites, tidal characteristics have been
derived at exospheric altitudes \citep{Forbes_etal09}. 

In the advent of the satellite technology, observations of internal waves
have dramatically increased, providing an unprecedented 
global view of the activity of these waves. 

\section{Techniques of Modeling Internal Wave Processes}
\label{sec:model}

Although observations suggest that wave-like structures are continuously 
present in the upper atmosphere \citep[e.g.,][]{Djuth_etal04, Livneh_etal07}, 
it is not always possible to determine their propagation and dissipation 
characteristics simultaneously and unambiguously. Theoretical, numerical, 
and global modeling studies allow for the analysis of various physical 
processes that influence the propagation of internal waves from their 
sources to regions where they strongly interact with atmospheric flow. 
Therefore, theoretical and numerical approaches are crucial for understanding
mechanisms of wave-flow interactions, and for better interpreting observations. 

\subsection{Theoretical Studies}
\label{sec:theory}
Theoretical (analytical) studies of internal waves have a long history.
They cover all aspects of wave dynamics -- generation, propagation, 
and dissipation, and provide their basic understandings. 
Although theoretical approaches are, generally, possible for highly 
simplified models, their results are widely used for parameterizing
subgrid processes unresolved by numerical models. A comprehensive 
review of these studies for gravity waves is given in the work of
\citet{FrittsAlexander03}. In this section we briefly summarize them 
focusing on recent developments.

Understanding and quantification of wave sources in the lower atmosphere 
seem to be the most challenging part to date. Overall, internal waves
are excited when air parcels experience vertical displacements in a 
stably stratified fluid. The main mechanisms of them in the real 
atmosphere include orography, convection, and jet/front systems. 
Flow over topography is the major source of atmospheric GWs with slow 
(with respect to the surface) horizontal phase speeds. 
These harmonics are especially important in the stratosphere, where 
they are responsible for ``hot spots" of wave activity that are tied 
up to particular geographical locations \citep{Hoffmann_etal13}.
These slow waves are easily filtered out by the mean wind, such that 
harmonics with larger phase velocities dominate in the middle atmosphere 
and thermosphere. Convective generation through various mechanisms, 
mostly in the tropics, is a source of fast and predominantly short GWs. 
Theoretical investigations that paved ways for parameterizing them in 
general circulation models are reviewed by \citet{Kim_etal03}.

GWs are ubiquitous in the middle and upper atmosphere, and are not
limited to low latitudes and/or particular hotspots. Tropospheric jet/front
systems are identified as the main source of these waves. Although 
vertical displacements of air parcels in jets and fronts are obviously 
responsible for generation, the exact mechanism through which GWs are
excited is not known, and is a subject of intense theoretical studies.
The main candidates are geostrophic adjustment, Lighthill radiation, various
instabilities and transient generation. Geostrophic adjustment implies a
presence of GWs superimposed on a slower and more ``balanced" flow. 
This leaves out the question of why they are present at first hand 
by simply stating that this is a fundamental property of the flow. 
Other theoretical mechanisms strive to explain this. Lighthill radiation 
mechanism of GWs is based on an analogy with the generation of acoustic 
waves by vortical turbulent motions. 
In the heart of it lies the nonlinearity of flows, by means of which 
energy from slower vortical modes is transferred to fast divergent
ones (gravity waves). The two mechanisms can probably be related if one
extends the geostrophic adjustment to include nonlinearity. With that,
a continuous competition occurs between the tendencies to distort the 
flow from a balanced state due to nonlinear advection, and to restore it 
by excitation of GWs. Remarkably, this approach yields results similar 
to the Lighthill expressions for wave sources \citep{MedvedevGavrilov95}. 
A conceptually close,
but analytically different approach is taken by proponents of the so-called
mechanism of ``spontaneous adjustment emission". One more line of theoretical
research is related to ``unbalanced" instabilities, that is, such conditions
under which balanced flows still remain stable, but the present infinitesimal
GWs become explosively unstable. Transient generation mechanism is
conceptually close to the unbalanced instabilities, but assumes that amplitudes
of thus generated waves can be predicted within the linear theory. An
excellent and insightful review of these theoretical developments has recently
been given by \citet{PlougonvenZhang14}.

\subsection{Numerical Modeling Techniques}
\label{sec:nummeth}
Numerical modeling allows for studying wave processes under more realistic
and complex conditions. Two major techniques include (1) ray tracing, and (2)
direct hydrodynamical simulations. Ray tracing is the least computationally
expensive among them. It calculates paths of narrow wave packets
centered around a harmonic with the given frequency and wavenumber through
regions
with varying propagation conditions. The information on wave phases is lost
under this approach, and the entire field can be represented as a collection
of a large number of wave packets and the corresponding very narrow beams
(rays). This technique has its limitations. First, it is applicable only if
the background varies slowly both spatially and temporally (compared to the
wavelength and period of a given harmonic, correspondingly), and, thus,
cannot be used for studying processes like wave break-up or nonlinear
interactions. Secondly, because no information on wave phases exists, processes
like diffraction and interference cannot be studied with the ray-tracing
techniques. Moreover, mathematical singularities called ``caustics" may
arise when ray paths of different packets come close. The advantage of the
ray tracing method is that trajectories are invariant in time, that is, it
can be applied to studying propagation of wave packets from their sources as
well as for identification of sources by tracking their paths back in time.
These approaches constitute direct and reverse ray tracing techniques,
respectively. They are widely employed for interpretation of observations and
linking wave signatures in the upper atmosphere to sources in the troposphere
\citep[e.g.,][]{Evan_etal12, Paulino_etal12, Pramitha_etal14}. A convenience of 
the ray tracing technique is that the same model can be used in both modes
\citep{VadasFritts09,Vadas_etal09}. More on the application of the ray-tracing
GW observations and implications for modeling can be found in the
paper of \citet{Ern_etal13}.

The method of direct wave simulation is based on numerical solution of the
fundamental (primitive) equations of hydrodynamics, and, therefore, does not
have many limitations of the ray tracing. It allows one to study various
aspects of wave propagation in the entire atmosphere, including the upper 
thermosphere, where molecular diffusion and ion friction substantially
alter the physics of waves. Normally, the hydrodynamic equations are
linearized with respect to the larger-scale mean flow (still retaining
nonlinear terms for disturbances). This formalism permits simulations of
propagation, refraction, ducting, critical layer filtering and dissipation 
of internal waves under a variety of realistic background distributions of 
wind and temperature \citep[e.g.,][]{LiuX_etal13, Yua_etal09}. In certain 
cases, direct wave models can be applied to studying wave propagation from 
particular sources like tsunami \citep{Occipinti_etal11}, or earthquakes
\citep{Matsumura_etal11}. Increasing computing power enables 
a consideration of weak nonlinear interactions between harmonics
\citep{Huang_etal14}, their break-ups \citep{GavrilovKshe13}, and even 
turbulence formation \citep{Fritts_etal09}. 
The main challenge with the direct wave simulation method comes from the 
fact that GWs may have fast phase speeds, integration requires a long 
time, and model domains must have significant sizes. This problem is 
usually circumvented by imposing periodic lateral boundary conditions, 
which prevent the accounting for
dispersion of wave packets. Extending domains of integration brings the
direct wave models closer to another class of numerical tools, namely to
general circulation models.

\subsection{General Circulation Modeling}
\label{sec:gcm}
General Circulation Models or Global Climate Models (GCMs) are 
three-dimensional (3-D) complex mathematical models that solve the 
fundamental equations of motion, energy, and continuity on a sphere. 
The numerical solutions of the conservation equations enable a 
simulation of the atmospheric dynamics, and an investigation of their 
interactions with the underlying physical, chemical, and radiative 
processes. The progress with digital computers in the second half of 
the 20th century have prompted the success of GCMs as research tools. 
The primitive equations are discretized and then solved numerically 
to simulate temporal evolution of atmospheric fields, such as wind, 
temperature, and density/pressure, under various boundary and external 
forcing conditions. In GCMs extending into the
ionosphere, conservation equations for the plasma are solved in addition 
to the neutrals \citep[e.g.,][]{GardnerSchunk11, Yigit_etal12a}. 
  
Depending on the regions of the atmosphere covered
by GCMs, they can either be global or limited-area ones. 
GCMs are also distinguished by the vertical layers they focus on:
lower/middle atmosphere-, upper atmosphere- and ``whole atmosphere" models. 
In general, lower/middle atmosphere models
typically extend from the ground up to the mesosphere or lower
thermosphere \citep[e.g.,][]{Manzini_etal06};
upper atmosphere models cover the thermosphere-ionosphere
from the mesopause to exobase \citep[e.g.,][]{GardnerSchunk11}. 
Whole atmosphere models are being increasingly developed, and, as
the name suggests, they perform calculations 
from the surface or lower atmosphere to the upper thermosphere
\citep[e.g.,][]{Liu_etal10b}. 
The horizontal and vertical resolutions combined with the assumed time 
step define the spatio-temporal capabilities of a GCM.

GCMs provide a variety of advantages. One of them is the ability to 
conduct control simulations. In the real atmosphere, physical processes 
occur simultaneously, and isolating individual processes is a 
challenging task. In a model, one can selectively turn on and off 
physical processes to determine their significance. Therefore, GCMs are 
useful tools for aiding the interpretation of observations. GCM output 
is easier to analyze because it contains all simulated fields, unlike 
with observations, which are always limited to certain parameters and 
incomplete.

However, no scientific tool comes without shortcomings. First, GCMs do 
not solve exactly the original partial differential equations, but
their algebraic approximations on a finite number of grid points, or 
elements, or spectral harmonics. Therefore, it is important that 
numerical methods are verified to ensure stability and convergence 
of numerical solutions to physical ones, and the model results must be 
validated. Because temporal and spatial resolutions of GCMs are limited, 
there are always scales of motions that cannot be properly resolved, 
and the subgrid-scale processes have to be accounted for, or 
``parameterized". For that, any type of physics that is not
self-consistently captured in GCMs should be mathematically described, 
ideally from first principles. Some examples of parameterizations in 
models are cloud microphysics, convection, eddy diffusion, and gravity waves.

Older as well as many current GCMs used relatively coarse horizontal 
grids, typically, a few degrees in longitude and latitude. 
Such resolutions are sufficient for modeling large-scale waves such 
as solar tides, planetary and Kelvin waves, however, they are inadequate 
for reproducing smaller-scale GWs. This prompted the development of GW 
parameterizations, among which the first were of \citet{Lindzen81} and
\citet{Matsuno82}. 
The progress in computational capabilities facilitates enhancing model 
resolutions, thus enabling GCMs to capture larger portions of small-scale 
GWs. For instance, \citet{Tomikawa_Etal12} have used a GCM with a T213 
spectral truncation, which correspond to a $0.5625^\circ$ longitude-latitude
resolution. 

Overall, state-of-the-art GCMs become increasingly more sophisticated and
complex in comparison with their predecessors. They include more physical
processes, higher resolution to capture dynamics at smaller scales, and can be
coupled together with other numerical models to form so-called ``climate system
models". An example of such model, are whole atmosphere GCMs extending from the
surface to the upper thermosphere \citep[e.g.,][]{Liu_etal10b, Liu_etal13}.

Development of GCMs requires extensive efforts typically by a group of
researchers. There are community models that are available to a broader
community. National Center for Atmospheric Research (NCAR) Community Climate
System Model is freely available to scientists worldwide. Extensive support is
provided to help guide users. This modeling framework is a product of
collaboration between researchers at NCAR and their national and international
collaborators. On the other hand, there are models that originate from a
specific research group. For example, the Ground-to-topside Model of Atmosphere
and Ionosphere for Aeronomy (GAIA) is a full atmosphere-ionosphere model that
has been developed at Kyushu University.

Some applications of numerical and GCM methods in the context of the
investigations of wave effects in the upper atmosphere are presented
in the next section.

\section{Wave Propagation and Consequences in the Upper Atmosphere}
\label{sec:effect}
Internal waves affect the momentum, energy, and composition balance of the
middle atmosphere through a variety of effects \citep[e.g., see reviews
by][]{FrittsAlexander03, Becker11}. Observational and modeling studies 
have shown that small-scale GWs \citep[e.g.,][]{Yigit_etal09} and solar
tides \citep[e.g.,][]{Oberheide_etal09} can directly propagate to the upper
atmosphere as well. We next focus on the upward propagation of these waves 
from the lower atmosphere to the thermosphere-ionosphere system, and the 
resulting effects. 

\subsection{Gravity Wave Effects}
\label{sec:gw}
Earlier studies extensively employed theoretical calculations and 
numerical simulations to characterize GW propagation and dissipation in the 
thermosphere-ionosphere \citep{Volland69, Hooke70,Klostermeyer72,HickeyCole88}. 
Recent numerical studies include more physical processes under more realistic
atmospheric conditions. In particular, the response of the
thermosphere-ionosphere to localized GW sources in the lower atmosphere have
been investigated. \citet{VadasLiu09} have considered the dissipation of 
GWs originated from a deep convective plume in Brazil. They found that the
resulting localized momentum deposition is the source of large-scale secondary
GWs and traveling ionospheric disturbances. The effects of GW dissipation 
in the thermosphere have recently been a subject of detailed studies
employing direct wave simulation models \citep{Hickey_etal09, Hickey_etal10b,
Walterscheid13, Heale_etal14}.

While theoretical studies and idealized numerical simulations can 
provide only a limited insight into gravity wave propagation and 
effects in the upper atmosphere, GCMs calculate four-dimensional geophysical
fields that can offer a more comprehensive view of the 
global atmosphere and coupling mechanisms therein. However, until 
recently, GW propagation to the thermosphere-ionosphere has been studied
with GCMs to a lesser extent. The reason for that is a combination of 
the following limitations: 
(1) Gravity waves that are capable of directly propagating from the lower
atmosphere into the upper atmosphere are rather small-scale and short-period, 
and cannot be captured to a large extent in GCMs; 
(2) Middle and upper atmosphere models were detached: the former
extended only up to the mesosphere \citep[e.g.,][]{BovilleRandel92, 
Beagley_etal97, Manzini_etal06}, while the latter had their
lower boundaries at around
80--90 km \citep[e.g.,][]{Roble_etal88, Richmond_etal92}; 
(3) GW parameterizations have primarily been designed for middle 
atmosphere models, and did not account for wave dissipation processes
appropriate for the atmosphere above the turbopause
\citep[e.g.,][]{AlexanderDunkerton99}. The deficiencies of the existing GW 
parameterizations were addressed by \citet{Yigit_etal08},
who have developed an ``extended nonlinear spectral" GW scheme
suitable for use in whole atmosphere models.

\begin{figure}\begin{center}
    \includegraphics*[width=1.0\textwidth,angle=0]
    {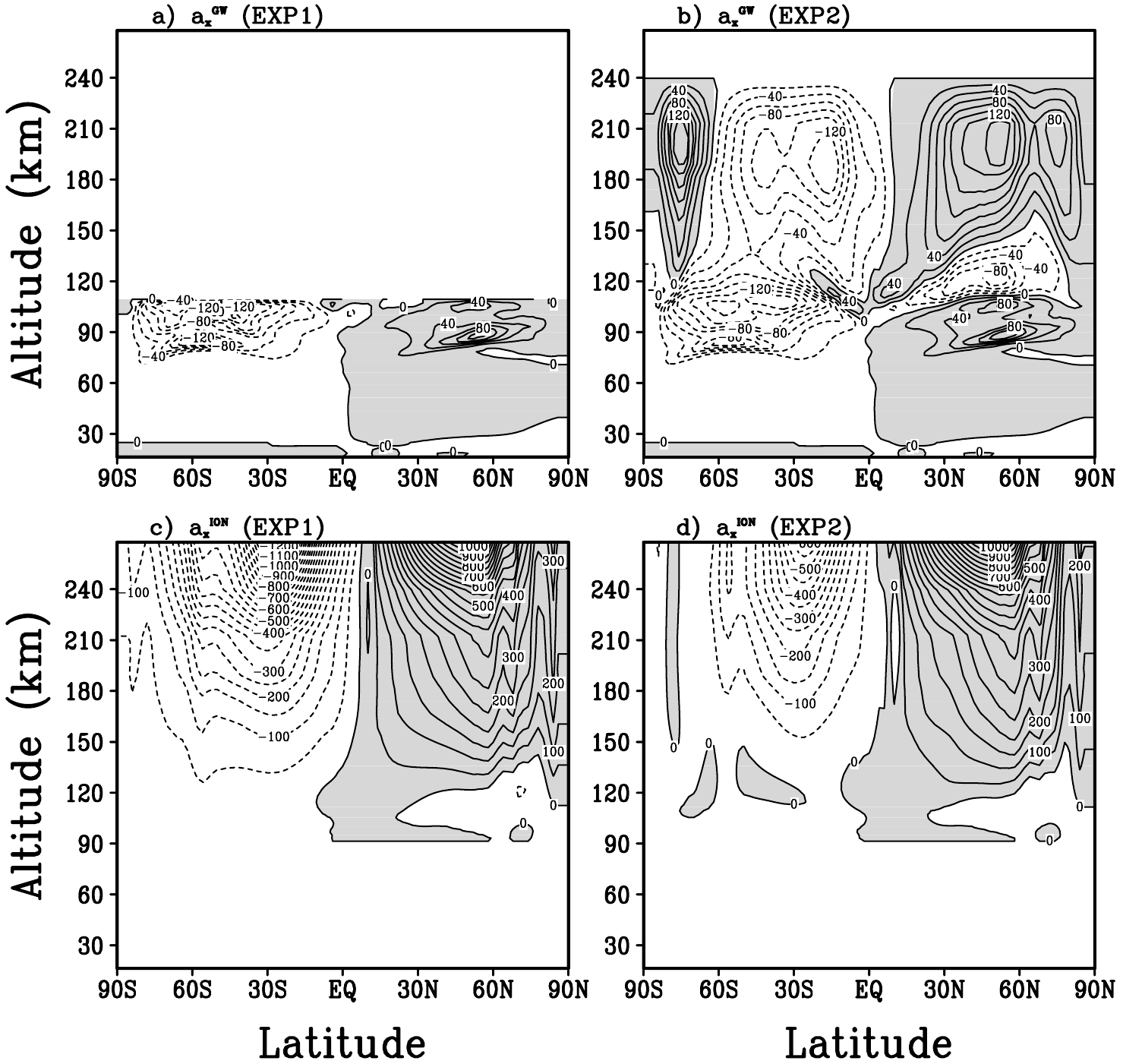}
  \end{center}
  \caption{Altitude-latitude cross sections of the zonal mean zonal GW drag
    (upper panels) and ion drag (lower panels) in m s$^{-1}$ day$^{-1}$ in
    the cut-off (EXP1, left) and extended simulations (EXP2, right) with the
    CMAT2 GCM \citep[Figure 3,][]{Yigit_etal09}.}
  \label{fig:gwmodel}
\end{figure}

For the first time, \citet{Yigit_etal09} have implemented the extended
parameterization into the Coupled Middle Atmosphere and Thermosphere-2 
GCM \citep[CMAT2,][]{Yigit09}, and simulated a global view of the 
small-scale GW propagation into the thermosphere. 
Figure~\ref{fig:gwmodel} compares the altitude-latitude cross-sections 
of the zonal mean zonal momentum deposition (``GW drag", upper row), 
and ion drag (lower row) during a solstice. In the first ``cut-off" 
simulation (EXP1), GW effects above the turbopause were neglected, which 
is similar to the use of conventional middle atmosphere parameterizations. 
The second simulation (EXP2) with the GW scheme turned on at all heights 
demonstrates that the subgrid-scale nonorographic GWs of the 
tropospheric origin are not only non-negligible in the thermosphere, but 
produce dynamical effects that are comparable to those by ion drag in 
the F region. 

\begin{figure}\begin{center}
    \includegraphics*[width=1.0\textwidth,angle=0]
    {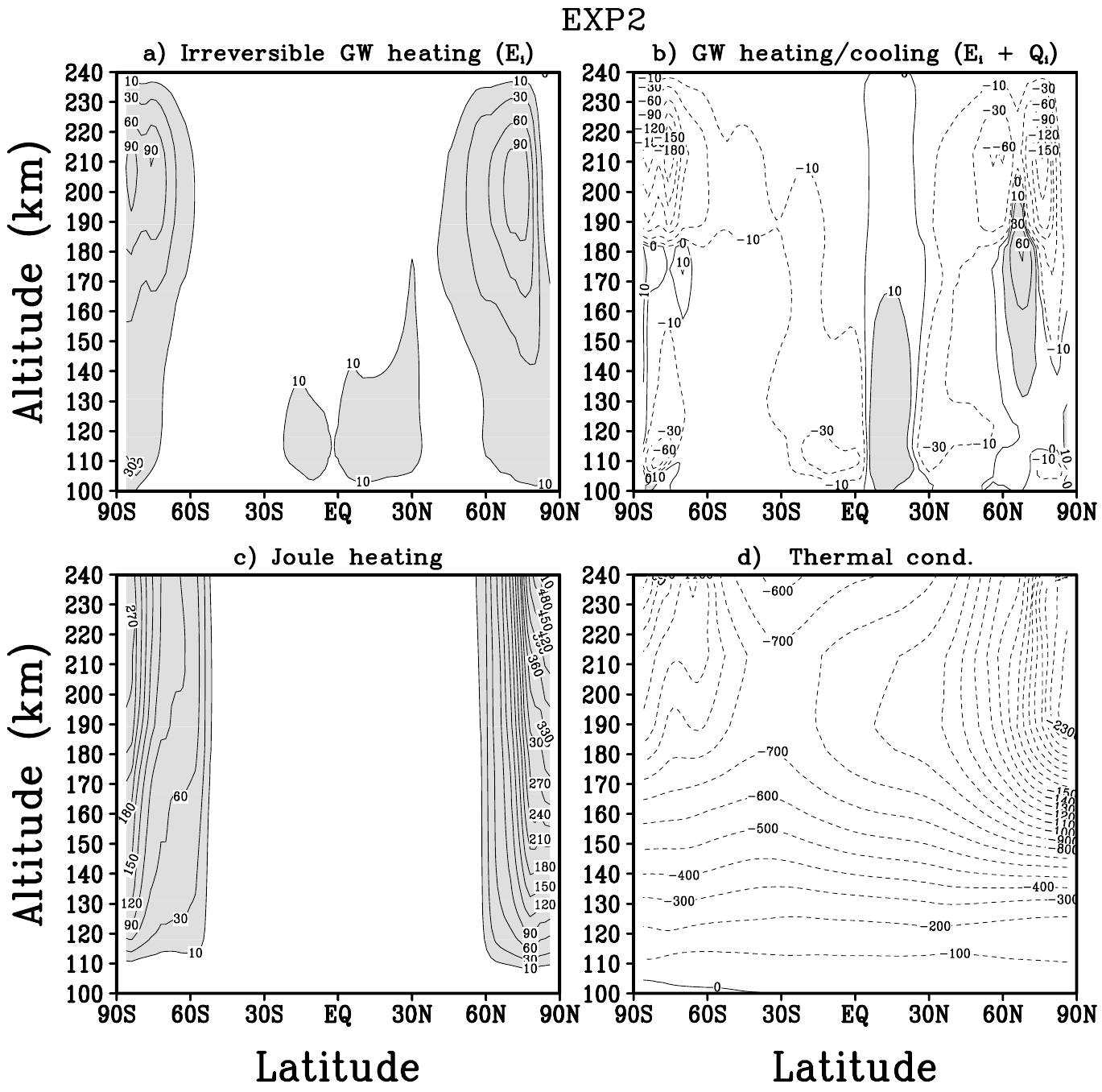}
  \end{center}
  \caption{Altitude-latitude cross sections of the mean zonal mean 
    a) GW irreversible heating,
    b) GW heating/cooling,
    c) ionospheric Joule heating, and
    d) cooling by thermal conduction 
    in K day$^{-1}$ for June solstice at low solar activity
    simulated by the CMAT2 GCM, implementing the extended nonlinear gravity
    scheme \citep[Figure 2,][]{YigitMedvedev09}.}
  \label{fig:gwheat}
\end{figure}

Thermal effects (heating and cooling rates) of lower atmospheric GWs in the
thermosphere are also significant.  Figure~\ref{fig:gwheat} shows the
altitude-latitude cross-sections of the calculated irreversible heating rates
due to dissipating GW harmonics, and the total heating/cooling rates (that
include the differential heating and cooling in addition to the irreversible
heating) from the work of \citet{YigitMedvedev09}. The former is comparable with
the Joule heating (Figure~\ref{fig:gwheat}c), and the latter with the cooling by
molecular thermal conduction (Figure~\ref{fig:gwheat}d).  Note that ion drag and
Joule heating are known to be two major dynamical and thermal processes in the
upper atmosphere \citep{Killeen87, Wilson_etal06, YigitRidley11a}. Thus, global
effects of GWs compete with the effects of ion-neutral coupling in the upper
atmosphere, and cannot be neglected.

\begin{figure}\centering
  \includegraphics*[width=1.0\textwidth,angle=0]
               {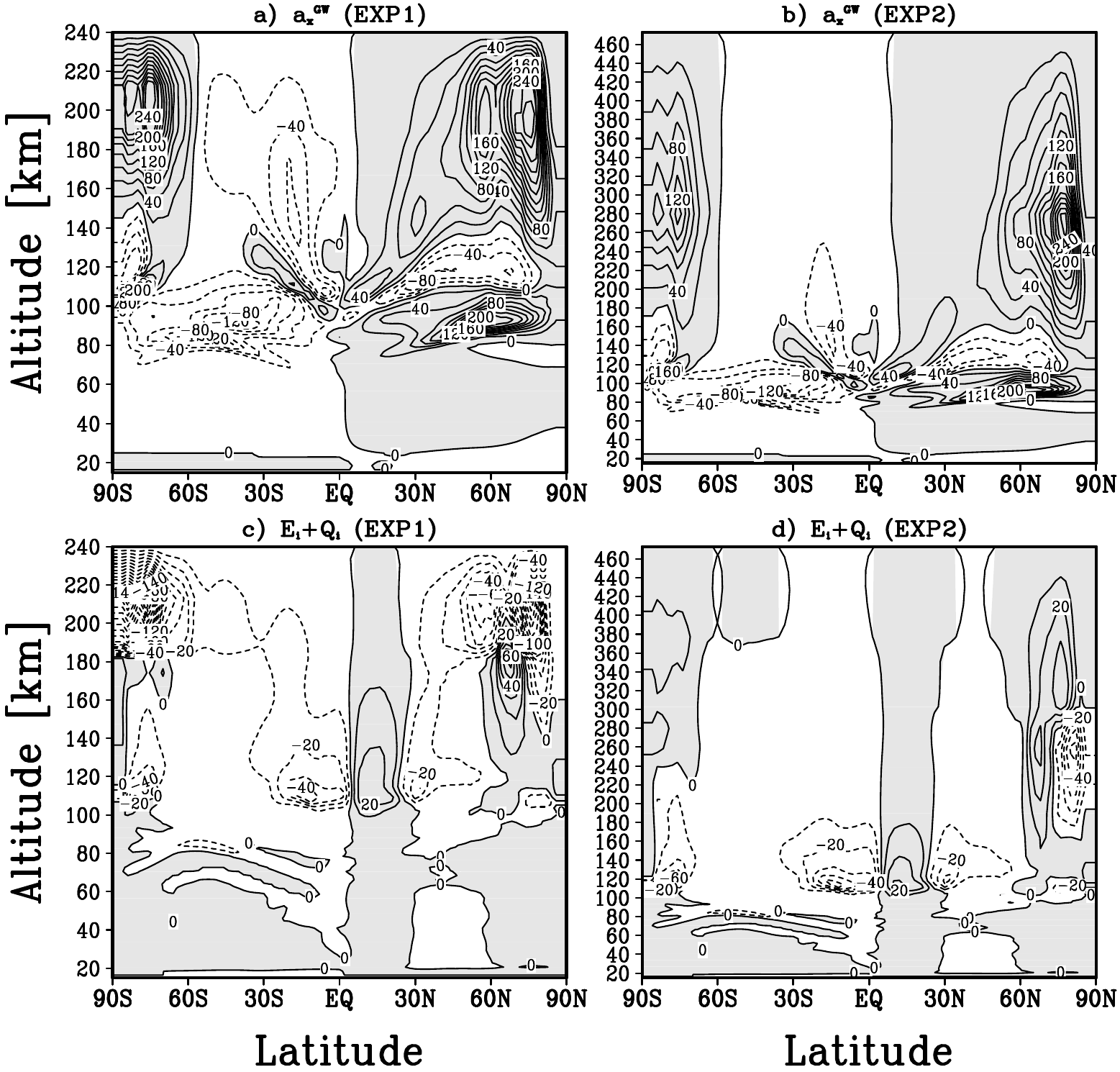}
  \caption{Altitude-latitude cross-sections of the mean zonal mean GW drag
      (panels a and b) and total GW heating/cooling (panels c and d) at low
      (left, EXP1) and high solar activity (right, EXP2) simulated with CMAT2
      using the extended nonlinear scheme \citep[Figure
        10,][]{YigitMedvedev10}.}
  \label{fig:gwsolar}
\end{figure}

Further GCM studies with the implemented extended GW scheme have provided more
insight into the propagation of GWs into the thermosphere during equinoctial
seasons \citep{Yigit_etal12b} as well as during periods of high solar activity
\citep{YigitMedvedev10}. Figure~\ref{fig:gwsolar} shows their results for the
mean zonal GW drag (panels a and b) and total GW heating/cooling (panels c and
d) at low (left, EXP1) and high solar activity (right, EXP2). Maximum
propagation altitude and thermospheric effects of lower atmospheric GWs are very
sensitive to solar activity. During high solar activity, GWs propagate to
altitudes of up to 450 km at high-latitudes, and produce mean effects of up to
240 m~s$^{-1}$~day$^{-1}$. At low solar activity, the mean effects are overall
larger, in particular, in the winter hemisphere, but the penetration of GWs into
the thermosphere is lower in altitude.

Higher resolution GCMs extending into the thermosphere have recently been
applied to simulate GW propagation and dissipation.  \citet{MiyoshiFujiwara08}
have used a spectral surface-to-exobase GCM with a T85 ($1.4^\circ\times
1.4^\circ$ longitude-latitude) resolution to examine GW characteristics in the
mesosphere and thermosphere. They found that a great portion of shorter-period
(and faster) GWs penetrate from the lower atmosphere to the heights of the F
layer.  More recently, \citet{Miyoshi_etal14} employed the same gravity
wave-resolving model to estimate the magnitudes and patterns of GW activity,
momentum deposition (``wave drag") in the thermosphere. Their results
confirmed those first obtained by \citet{Yigit_etal09} and
\citet{Yigit_etal12b} for the solstitial and equinoctial conditions.  These
results provided conclusive evidences not only of the dynamical importance of
GWs propagating to the thermosphere from below, but also that their
thermospheric effects can be successfully captured by parameterizations in
lower-resolution GCMs. Such GWs affect the ionosphere as well.  Recently,
\citet{Shume_etal14}'s observational studies indicated that GWs forcing could
have been responsible for short-period electrojet oscillations observed over
Brazil.

Direct effects of the lower atmospheric acoustic-gravity waves on the upper
atmosphere are observed distinctly during earthquakes/tsunamis as well
\citep[e.g.,][]{HekiPing05}. For example, the investigation of the Sumatra and
Tohoku-Oki tsunamis have revealed detailed dynamical effects of GWs, which had
not been anticipated before \citep[e.g.,][]{Makela_etal11, Roland_etal11}.

\subsection{Tidal Effects}
\label{sec:st}
Although, a significant portion of the tidal energy is absorbed in the lower
thermosphere, further propagation of tidal signatures occur beyond the
turbopause. Tides of lower atmospheric origin can be observed in the
thermosphere by satellites \citep{OberheideForbes08b}, and their effects on the
thermospheric composition have been derived, for example, from the data obtained
with the TIMED and SNOE satellites \citep{OberheideForbes08a}.  A number of
researchers have provided evidences for tidal modulation of the low-latitude
thermosphere \citep[e.g.,][]{Luhr_etal07, Forbes_etal09, Liu_etal09,
  Oberheide_etal09}. \citet{Kwak_etal12} have identified the signatures of the
wavenumber-three eastward travelling nonmigating diurnal tide (DE3) in the
thermosphere, and concluded that they are a persistent feature of the
thermosphere during low solar activity.

\begin{figure}\centering
  \includegraphics*[width=1.0\textwidth,angle=0]
  {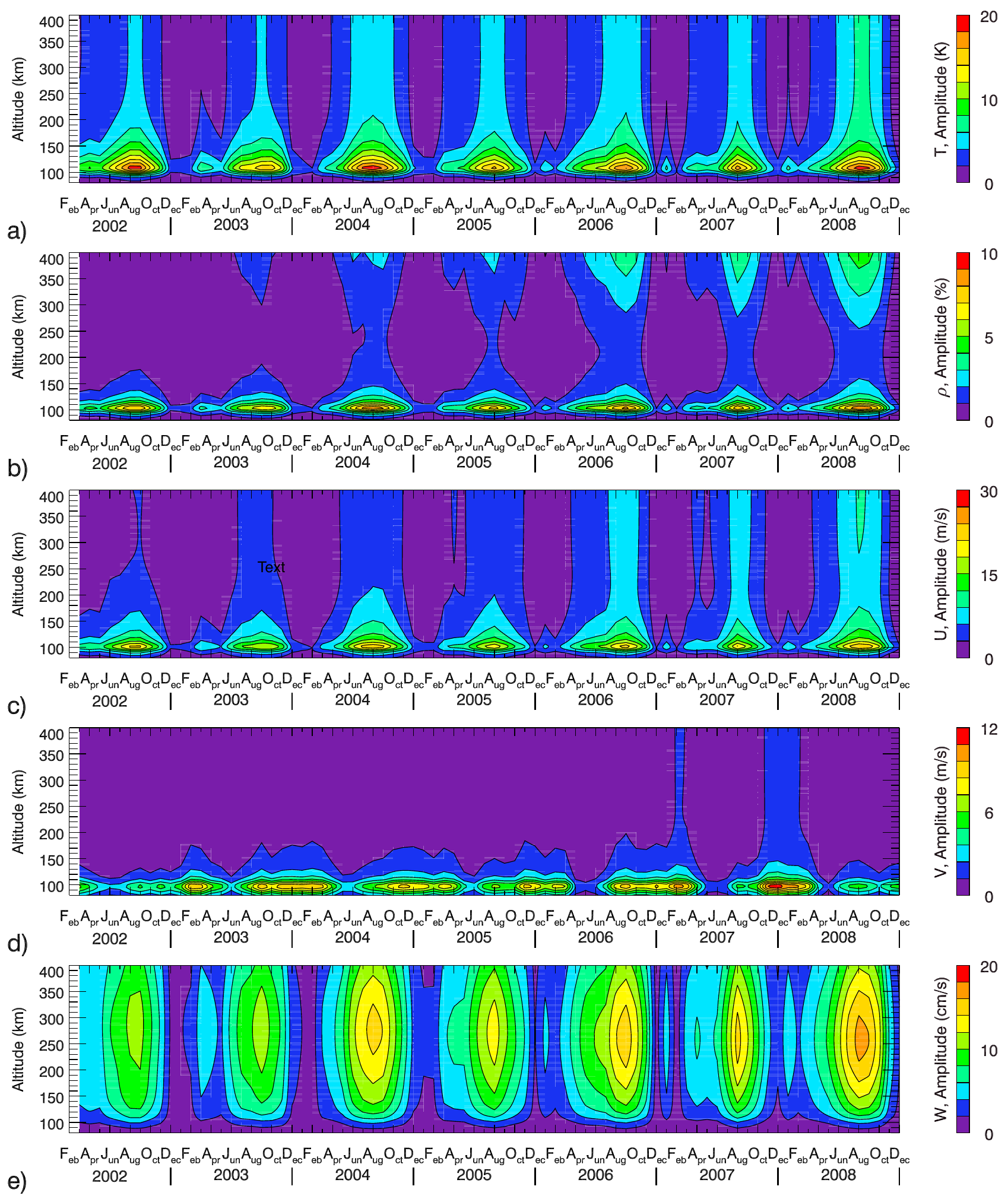}
  \caption{DE3 tidal propagation into the thermosphere in terms of the tidal
    amplitudes for the temperature, density, zonal wind, meridional wind, and
    vertical wind  \citep[Figure 13,][]{Oberheide_etal09}.}
  \label{fig:tide}
\end{figure}

\citet{Oberheide_etal09} have studied the question of how much of the tidal 
signatures propagates directly to the upper 
atmosphere. They have analyzed data from the TIMED in the MLT, and CHAMP 
satellite at $\sim$400 km, using the Hough Mode Extension technique. 
Figure~\ref{fig:tide} from their work demonstrates a direct propagation of
the DE3 tide from the MLT to the upper thermosphere in
terms of the vertical amplitude distribution for the tidal disturbances of the
field variables $(T,\rho,u,v,w)$. The panels a--e show that the DE3 variations
of temperature extend higher into the upper thermosphere than those of other
fields  in $(\rho,u,w)$, although the density 
fluctuations have a secondary peak around 400 km. 

Global effects of lower atmospheric tides can be readily investigated 
with whole atmosphere models. Their impacts can be estimated by turning on and
off the tidal activity in the lower atmosphere \citep{Yamazaki_etal13}.
Using a GCM extending from the ground to the exobase, \citet{Miyoshi_etal09} 
have shown that the solar terminator wave observed by CHAMP 
in the thermosphere is generated mainly by the superposition of upward
propagating migrating tides with wavenumbers 4--6. \citet{Jin_etal11} employed a
whole atmosphere-ionosphere GCM to investigate the relationship 
between the wavenumber four structure and upward propagation of the nonmigrating
tides. Global response of the ionosphere to the upward propagating tides from
below has been investigated by \citet{Pancheva_etal12} using the GAIA GCM along 
with the COSMIC observations. They have determined three altitude regions of 
enhanced electron density in the thermosphere-ionosphere, and
discovered the evidence that the wavenumber four ionospheric
longitudinal structure is not solely generated by DE3 tide.

\section{Vertical Coupling during Sudden Stratospheric Warmings}
\label{sec:ssw}

In this section, we focus on the observed and modeled effects of
sudden stratospheric warmings (SSWs) on the upper atmosphere. SSWs
are spectacular transient events in the winter Northern Hemisphere (NH) 
first discovered by \citet{Scherhag52}. The winter polar temperature 
dramatically increases within a few days following the breakdown 
or weakening of the stratospheric polar vortex as a consequence of 
planetary wave amplification and breaking. Such warmings are accompanied by
deceleration, and even reversals of the westerly zonal mean zonal winds at 10
hPa ($\sim 30$ km). \citet{Matsuno71} was the first to demonstrate 
with a simple dynamical numerical model that planetary waves and
their interactions with the zonal mean flow are responsible for SSWs. 
Further numerical studies confirmed \citet{Matsuno71}'s conclusion 
qualitatively \citep{Holton76, Palmer81}. \citet{Schoeberl78} provides one of
the earliest reviews of the theory and observation of
stratospheric warmings focusing on the middle atmosphere.

\begin{figure}\centering
  \includegraphics*[width=0.8\textwidth]
  {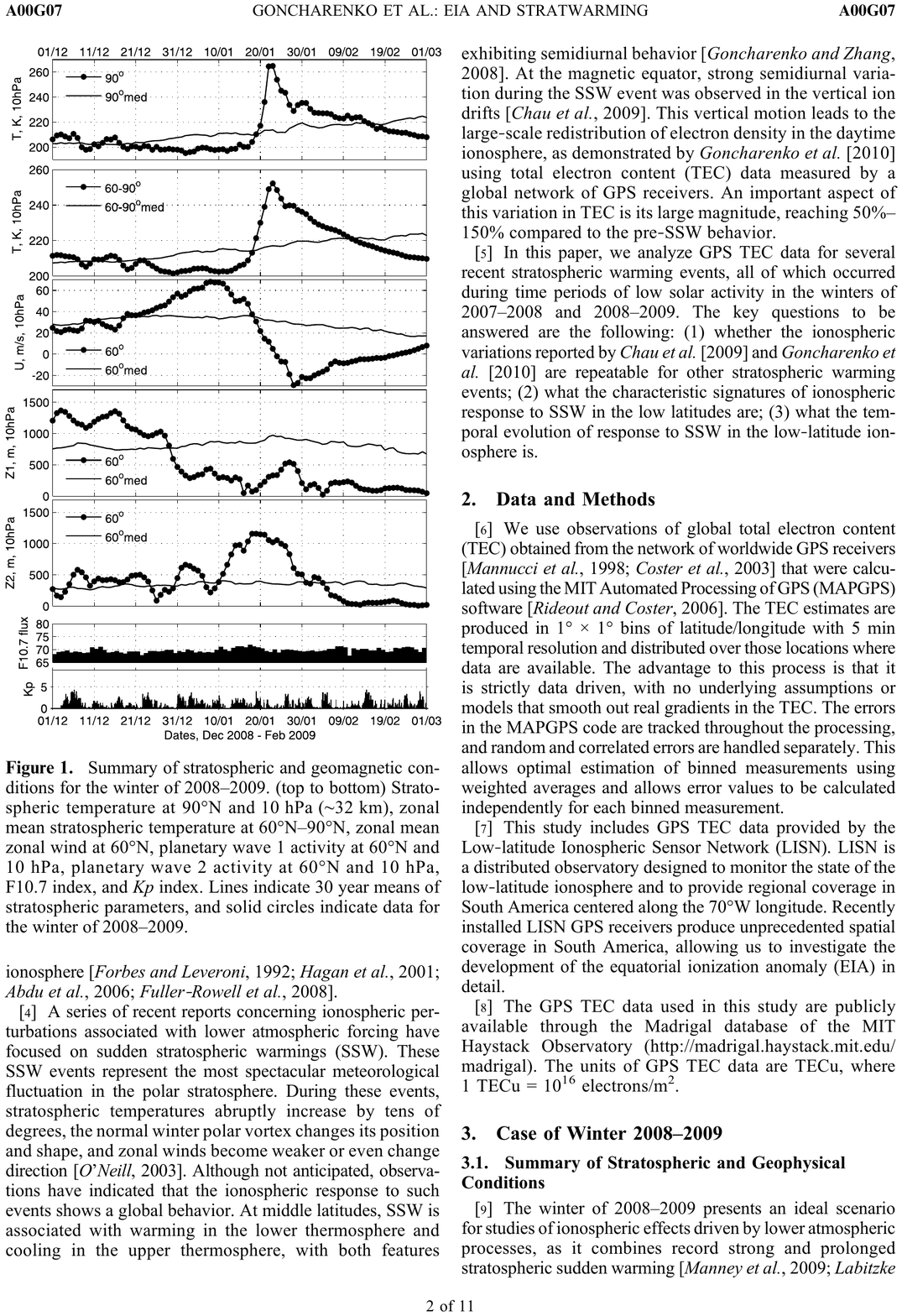}
  \caption{Stratospheric conditions for the winter of 2008--2009. Temperature at
    90$^\circ$ N and 10 hPa ($\sim$32 km), zonal mean temperature at
    60$^\circ$--90$^\circ$N , zonal mean zonal wind at 60$^\circ$N, planetary
    wave 1 and 2 activity at 60$^\circ$N and 10 hPa.  Solid lines indicate
    30-year means and solid circles indicate data for the winter of 2008--2009
    \citep[Figure 1,][]{Goncharenko_etal10}.}
  \label{fig:ssw}
\end{figure}

An observation of stratospheric conditions at 10 hPa during the major 
SSW that took place in the winter of 2008--2009 is shown in
Figure \ref{fig:ssw} adopted from the work by \citet{Goncharenko_etal10}. 
At 10 hPa, the zonal mean temperature at the Northern winter Pole 
increases from 200 K to more than 260 K within a few days. 
This warming is accompanied by a reversal of the zonal mean winds from 
westerly ($>60$ m~s$^{-1}$) to easterly ($< -20$ m~s$^{-1}$) at the same 
altitude, and is defined as a ``major" warming. If the winter pole warms 
up significantly, and the zonal mean zonal jet weakens, but does not 
reverse, the event is said to be a ``minor" warming.

Sudden changes of the morphology of the troposphere-stratosphere
during SSWs  \citep{Limpasuvan_etal04}, and the accompanying effects at 
higher altitudes provide a natural laboratory where atmospheric 
vertical coupling processes can be investigated. An increasing number 
of global observations indicate that SSW effects can be detected
beyond the stratosphere, that is, in the mesosphere-thermosphere-ionosphere. 
Strong upper atmosphere effects have been detected, in particular, 
during quiet magnetospheric conditions.
Using Millstone Hill incoherent radar data on ion
temperatures, warming in the lower thermosphere and cooling
above 150 km were observed during a minor SSW by \citet{GoncharenkoZhang08}.
\citet{Chau_etal09} observed a significant 
amount of semidiurnal variations in the ${\bf E}\times{\bf B}$ vertical 
ion drifts in the equatorial ionosphere during the winter 2007-2008 minor 
warming. \citet{Goncharenko_etal10} have investigated the impact of the 
2008-2009 major warming on the Equatorial Ionization Anomaly (EIA). 
Their analysis of GPS data at low-latitudes showed that a few days before 
the onset of the warming, appreciable local time variations were present 
in the magnitude of EIA. \citet{PedatellaForbes10} observed significant 
enhancement of the nonmigrating semidiurnal westward propagating tide 
with the zonal wavenumber one (SW1) during the 2009 SSW. 
Observational evidence from satellite measurements for the dynamical
coupling between the lower and upper atmosphere during SSWs have been
provided by \citet{Funke_etal10}. \citet{PanchevaMukhtarov11} have
found a systematic negative response of ionospheric plasma parameters 
($f_0F_2$, $h_mF_2$, and $n_e$) to an SSW, in the COSMIC
(Constellation Observing System for Meteorology Ionosphere and Climate) data. 

More recently, \citet{Goncharenko_etal13} have investigated the day-to-day
variability of the midlatitude ionosphere during the major SSW of 2010, and
discussed the occurrences of various wave structures in the upper atmosphere
during the warming.  They found enhanced semidiurnal and terdiurnal variations,
and raised the question of how these signals can propagate from the stratosphere
to the thermosphere. \citet{Kurihara_etal10} observed significant short-term
variations during a major SSW in the lower thermospheric zonal wind and
temperature retrieved from the EISCAT UHF radar in high-latitudes. Motivated by
the previous observational findings, modeling efforts of \citet{Liu_etal13}
demonstrated an appreciable local time and height dependence of the upper
atmospheric response to SSWs. Recently, SSW effects on the upper atmosphere
  are being increasingly studied in the Southern Hemisphere as well
  \citep[e.g,][]{Jonah_etal14}.

\begin{figure}\centering
  \includegraphics*[width=0.6\textwidth,angle=0]
  {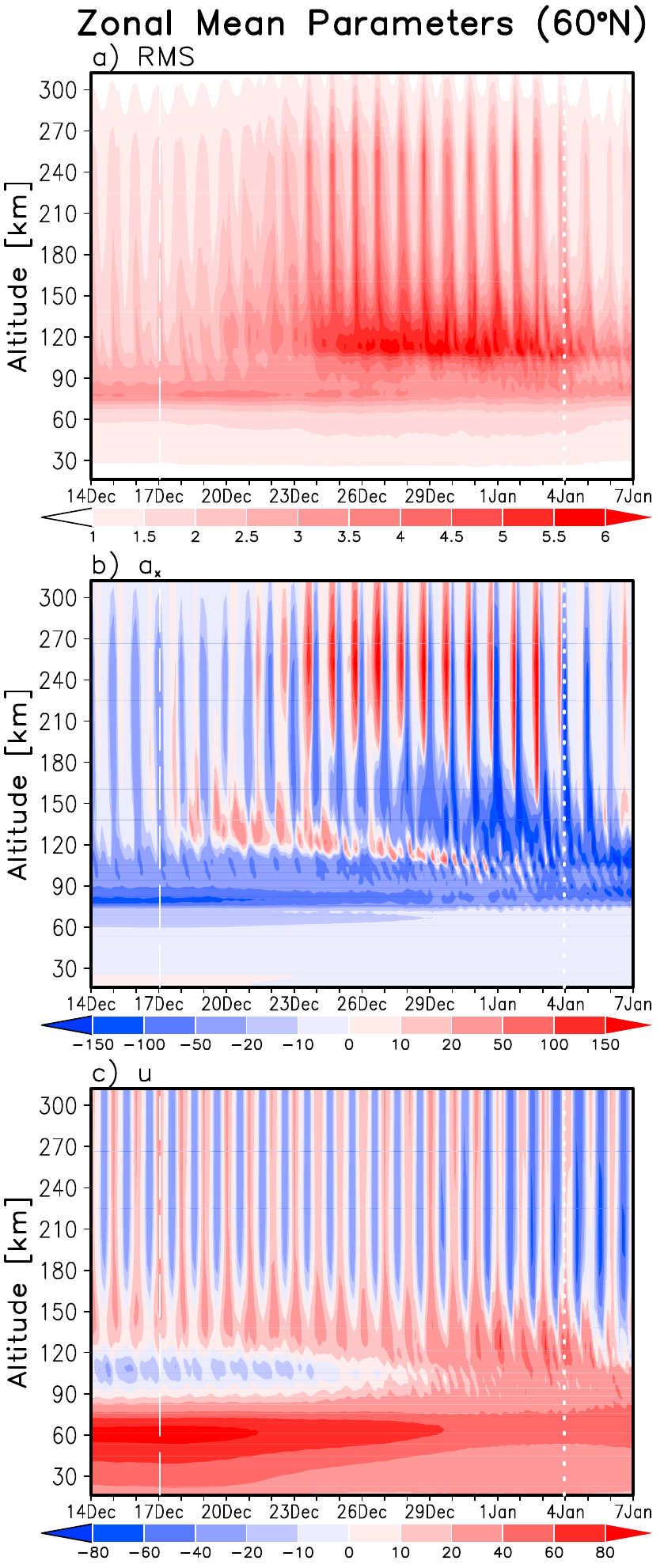}
  \caption{GCM simulation of the altitude-universal time variations of the
      zonal mean 
      a) root-mean-square zonal wind (RMS, GW activity) in m~s$^{-1}$,
      b) GW drag in m~s$^{-1}$~day$^{-1}$, and 
      c) large-scale zonal wind m~s$^{-1}$, during a minor warming
      \citep[Figure 2,][]{YigitMedvedev12}.}
  \label{fig:gwssw}
\end{figure}

Dramatic changes in the zonal mean zonal winds $\bar{u}$ in the stratosphere 
have a great impact on the propagation and dissipation of GWs, primarily
due to the alteration of the GW intrinsic phase speed, $c-\bar{u}$. 
The evolution of large-scale wind field $\bar{u}$ during SSWs can 
successfully be reproduced by GCMs 
\citep[e.g.,][and references therein]{CharltonPolvani07,delaTorre_etal12},
thus, providing an opportunity for establishing a link between
SSWs and variations of GW activity at altitudes up to the lower thermosphere
\citep{LiuRoble02, Yamashita_etal10a}. Applying this modeling approach,
\citet{YigitMedvedev12} investigated for the first time the global 
propagation of subgrid-scale GWs from the lower atmosphere to the thermosphere
above the turbopause during a minor warming. Figure \ref{fig:gwssw} 
shows the altitude-universal time distributions of the zonally
averaged a) GW activity, b) GW drag, and c) large-scale zonal wind. 
The two white vertical lines denote the SSW period, over which these
quantities experience significant changes. An enhanced propagation into the
thermosphere  causes an amplification of the eastward 
GW momentum deposition in the lower and upper thermosphere by up to a
factor of 6, which, in turn, affects the zonal mean wind. 

\begin{figure}\centering
    \includegraphics*[width=1.0\textwidth,angle=0]
    {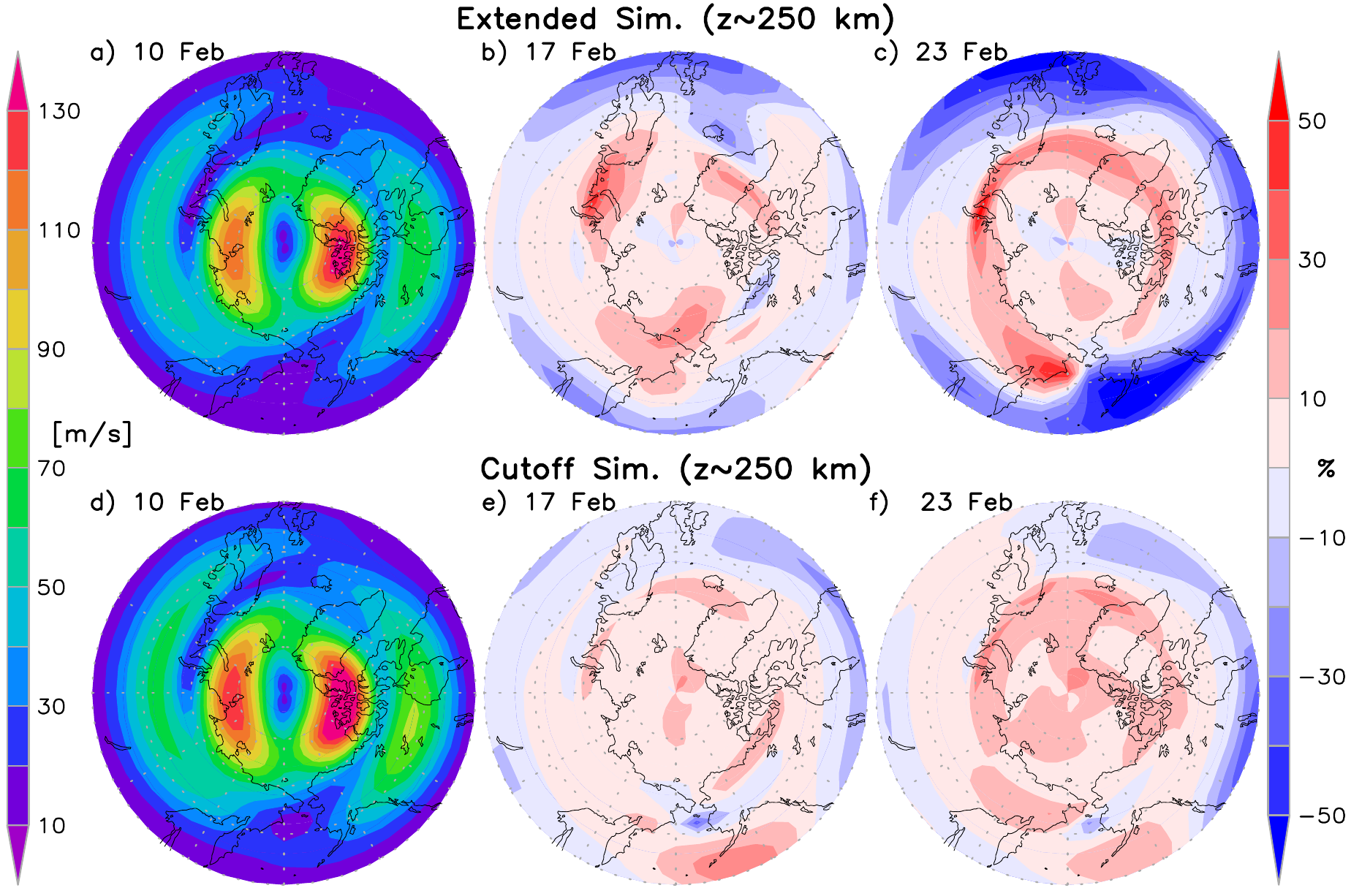}
  \caption{Northern Hemisphere polar stereographic projections at 250 km
      of small-scale
      variability on
      10 Feb and the percentage changes (second and third columns from left)
      during the later phases of the warming in the extended (upper row) and the
      cut-off simulations (lower panels) \citep[Figure 4,][]{Yigit_etal14}.}
  \label{fig:gwsswv}
\end{figure}

As more small-scale GW harmonics propagating from below reach the 
thermosphere during SSW events, they strongly impinge on the larger-scale 
flow upon their breaking and/or saturation. \citet{Yigit_etal14} have
investigated the GW-induced small-scale variability in the high-latitude
thermosphere over the life cycle of a minor SSW. Figure~\ref{fig:gwsswv}
presents the short-period (excluding tides) temporal variability of the
simulated zonal wind at Northern Hemisphere high-latitudes at 250 km 
during the different phases of the warming. 
The upper panel shows the results of the simulation, in which GWs were allowed
to propagate all the way up into the thermosphere, and the lower panel is for
the control simulation with GW propagation terminated above the turbopause. 
It is seen that GWs from below are responsible for a $\pm50\%$
change in the small-scale variability of the resolved zonal wind.

\section{Upper Atmosphere Variability}
\label{sec:uav}

The term ``variability" implies an existence of a mean state $\bar{s}$ 
with respect to which deviations s$^\prime$ are studied. 
Since any field $s$ is the superposition of the mean and deviations, 
$ s = \bar{s} + s^\prime$, the choice of the mean $\bar{s}$
determines the spatio-temporal structure of the variability $s^\prime$.
The practical importance of variability is that it is a source of 
uncertainty in the prognosis of the atmosphere-ionosphere.
The upper atmosphere is a highly variable region at all temporal and 
spatial scales ranging from minutes to decades, and from local to 
global scales. There are numerous technical challenges in studying
natural variability. It is not always easy to fully capture this 
variability due to observational and numerical constraints. Global models
always have a limited resolution. Distinguishing between a physical and
non-physical variability could be an observational challenge.
Because of that, and the fact that GW signatures often have no
well-defined wave-like structures, the variability can be studied as an 
additional characteristic of the wave field.

One of the spectacular upper atmosphere features is the variability 
of thermospheric vertical winds. Large vertical motions 
are continuously present in the upper atmosphere 
\citep{Price_etal95, Ishii_etal01} with appreciable variations
\citep{InnisConde01}. GCM studies have indicated that
nonhydrostatic effects are crucial in the spatio-temporal variability of
vertical winds \citep{YigitRidley11b, Yigit_etal12a}. 

Overall, the variability is observed not only in neutral winds 
and temperature, but also in composition \citep{Kil_etal11}, ion flows 
\citep{Bristow08}, electric fields \citep{Kozelov_etal08}, and Joule 
heating \citep{Rodger_etal01}. 
Besides the inherent variability due primarily to the nonlinearity of 
the underlying dynamics 
and physics, several sources external to the thermosphere-ionosphere
have been identified. They are changes in (1) the solar irradiation, 
(2) magnetospheric forcing, and (3) the lower atmosphere.
The first two are typically designated as ``space weather influences" 
from above, while the latter is the ``meteorological forcing" from below.
Despite the growing amount of observational data, separating 
the contributions of each mechanism to the overall variability is quite 
a challenging task. Concerning the subject of this review, one can 
state that studies of the variability due to the dynamical coupling 
from below are still at their infancy, although the potential of the 
lower atmosphere to contribute to the observed upper atmosphere 
variability has already been recognized \citep{Rishbeth06}.

How can the lower atmosphere influence the upper atmosphere variability?
There are a few pathways/physical mechanisms through which the lower
atmospheric variability imprints on the thermosphere-ionosphere system. 
One of them is the direct penetration of highly irregular GWs from 
below. Observations have revealed a continuing presence and persistence of
such waves, some of which have been discussed in this paper,
but more was given in the review paper of \citet{FrittsLund11}.
Due to the enhanced dissipation by molecular diffusion of harmonics with
shorter vertical wavelengths, mostly fast harmonics with longer vertical
scales can survive the propagation to the upper thermosphere. They can often
manifest themselves as traveling ionospheric disturbances (TIDs)
\citep[e.g.,][]{FujiwaraMiyoshi09}. Planetary and ultra-fast Kelvin wave
signatures are also observed in the thermosphere, although their vertical
extent to the upper thermosphere is strongly terminated by dissipation
\citep{Chang_etal10,Abreu_etal14}.

Another mechanism of small-scale variability is associated with GW breaking,
which occurs at scales much smaller than the wavelength. 
Localized events not only permeate the flow, but also give rise to 
short-period and long (fast) waves. Such mechanism of secondary 
excitation has been extensively studied 
\citep[e.g.,][]{Vadas_etal03,ChunKim08}, and found to be a likely source
of harmonics that can effectively propagate into the upper thermosphere
\citep{VadasLiu11}. Modulation of gravity wave propagation in the middle
atmosphere, e.g., during sudden stratospheric events \citep{Yigit_etal14}, 
or by enhanced dissipation during periods of increased solar
activity \citep{YigitMedvedev10} can influence the upper atmosphere
variability. GWs can influence the degree 
of ion-neutral coupling primarily by modulating ion-neutral 
differential velocities. Depending on the degree of GW 
penetration into the thermosphere and the plasma flow patterns, such 
modulating can constitute a significant source of variability
\citep{Yigit_etal14}.

The third pathway does not require a direct wave propagation to the upper
atmosphere, but involves a chain of additional physical mechanisms to imprint
lower atmospheric inhomogeneities onto the upper layers. A notable example is
the wavenumber-four longitudinal structure of the low-latitude ionosphere seen
in electron density and temperature, nitric oxide density, and F-region neutral
winds \citep[see][for more observational evidences]{Ren_etal10}. It was
suggested in the work by \citet{Immel_etal06b} that the nonmigrating diurnal
tide with the wavenumber-three traveling eastward is the main source of the
wavenumber-four structure. The DE3 tide is generated by the latent heat release
in the tropical lower atmosphere, propagates to the MLT heights, where it
reaches significant amplitudes, and can be a dominant mode of the diurnal tide
during certain times \citep{Oberheide_etal11}. \citet{Immel_etal06b} suggested
that the DE3 tide modulates the ionospheric dynamo at the E-region, thus
affecting electric fields in the F-region along magnetic lines, and drives the
ionospheric wavenumber-four structure. GCM simulations \citep{Hagan_etal07} have
confirmed this mechanism, while the subsequent studies investigated its various
aspects \citep{Ren_etal10}.


\section{Open Questions and Concluding Remarks}
\label{sec:conc}
A concise review of vertical coupling in the atmosphere-ionosphere system has
been presented here, focusing on the role of internal waves as the main
vertical coupling mechanism.  Considerable progress has been made, over the
past decade, in the appreciation of the role, which these waves play in the
dynamical coupling between the lower and upper atmosphere.  Internal waves
include planetary Rossby and Kelvin waves, tides, and gravity waves. Due to
their ability to propagate vertically, internal waves represent a dynamical link
between atmospheric layers.

There are many open questions that still remain in this research area, 
some of which are listed below. We do not intend to compile a full
list of them, but name the most basic and pressing, in our view,
unresolved problems.
\begin{enumerate}
\item What are the momentum fluxes and spectra of internal gravity waves
  penetrating into the thermosphere from below?
\item Can the sources of gravity waves be parameterized in terms of large-scale
  fields, such that the generation can be modeled by GCMs self-consistently,
  rather than introduced as external tuning parameters?
\item To what degree do the external energy sources (solar irradiation,
  geomagnetic activity) and the associated variability in the upper atmosphere
  affect the middle, and even lower atmosphere? What are the dynamical
  mechanisms?
\item What information from the lower atmosphere is needed to predict the
  dynamical variability above?
\end{enumerate}

Although the focus of this review has been on the Earth
atmosphere, internal wave coupling has wider implications, for instance, for
understanding circulations of other planets, like Mars \citep{myhb11},
and Venus \citep{Garcia_etal09}, as well.\\

\noindent
{\bf Acknowledgements}

The work was partially supported by German Science Foundation (DFG) grant
ME2752/3-1. Erdal Yi\u{g}it was partially supported by NASA grant
NNX13AO36G. The authors are grateful to Art Poland at George Mason University's
Space Weather Laboratory for his valuable comments on the manuscript.

\newpage
\begin{table}\centering
  \caption{Typical temporal scales of internal waves in the terrestrial
    atmosphere.}
  \begin{tabular}{lll}
    \hline
    Internal wave   & & Typical range of temporal scales \\
    \hline
    \hline
    Gravity wave    &  & few minutes to several hours ($2\pi/f,
      f=2\Omega\sin\phi$) \\
    Solar tide      &  & 1, 1/2, 1/3 days\\
    Planetary wave  &  & 2 to few tens of days\\
    Kelvin wave     &  & 3 to 20 days\\
    \hline
  \end{tabular}
  \label{tb:table1}
\end{table}

\end{document}